\documentclass[twocolumn,preprintnumbers,showpacs,amsmath,amssymb,amsthm,aps,prd,nofootinbib]{revtex4-1}
\usepackage{graphicx}
\usepackage{color}
\usepackage{amsthm}
\usepackage[utf8]{inputenc}
\usepackage{amsmath}
\usepackage{mathrsfs}
\usepackage[colorlinks,hyperindex]{hyperref}  
\usepackage{accents}
\usepackage{enumitem}
\usepackage[normalem]{ulem}
\definecolor{byzantine}{rgb}{0.74, 0.2, 0.64}
\definecolor{vividviolet}{rgb}{0.62, 0.0, 1.0}
\hypersetup
 {
   colorlinks,%
   citecolor=blue,%
   linkcolor=blue,%
   urlcolor=blue,%
 }
 
\def\setR{\mathbb{R}}

\newcommand{\sss}[1]{\scriptscriptstyle #1} 

\newcommand{\pbundle}[4]{#1(#2,#3,#4)} 
\newcommand{\sbundle}[3]{#1(#2,#3)} 

\newcommand{\Miv}[1]{{\color{red}{#1}}}

\newcommand{\Old}[1]{{\color{byzantine}{#1}}}

\begin{document}

\title[Cartan TEGR review]{Cartan approach to Teleparallel Equivalent to General Relativity: a review}

\author{ E.~Huguet$^1$, M.~Le~Delliou$^{2,3}$, and M.~Fontanini$^1$}
\affiliation{$1$ - Universit\'e de Paris, APC-Astroparticule et Cosmologie (UMR-CNRS 7164), 
 F-75006 Paris, France.}
\email{michele.fontanini@gmail.com\\
huguet@apc.univ-paris7.fr}
\affiliation{$2$ - Institute of Theoretical Physics, School of Physical Science and Technology, Lanzhou University,
No.222, South Tianshui Road, Lanzhou, Gansu 730000, P R China
}
\affiliation{$3$ - Instituto de Astrof\'isica e Ci\^encias do Espa\c co, Universidade de Lisboa,
Faculdade de Ci\^encias, Ed.~C8, Campo Grande, 1769-016 Lisboa, Portugal}
\email{(delliou@lzu.edu.cn,)morgan.ledelliou.ift@gmail.com}

\date{\today}
\begin{abstract}

In previous works, questioning the mathematical nature of 
the connection 
in the translations gauge theory formulation of Teleparallel Equivalent to General Relativity (TEGR) Theory led us to propose a new formulation using a Cartan connection. In this review, we summarize the presentation  of that proposal and discuss it from  a gauge theoretic perspective.

\end{abstract}
\maketitle
\tableofcontents
\section{Introduction}\label{SEC-Intro}

As a description of gravity, the Teleparallel Equivalent to General Relativity (TEGR) offers a symmetric and classically equivalent way of presenting the physics described by General Relativity (GR): the main difference in perspective being that GR uses curvature of spacetime to describe the effects of gravity requiring zero torsion, while TEGR requires zero spacetime curvature and fully encodes gravity in the torsional part.
This mirror approach to the description of gravity is not proprietary to TEGR, Symmetric Teleparallel Gravity (STGR), introduced in Ref.~\cite{Nester:1998mp}, is another example of a possible equivalent description where curvature and torsion are symmetrically taken to be null and non-metricity contains all the relevant physics.
In the many years TEGR has been around, its formal presentation has taken various forms
, from the translation-gauge approach of Ref.~\cite{Aldrovandi:2013wha}, the ``pure tetrads formalism" point of view in Ref.~\cite{Maluf:2011kf}, to the tensorial formalism of Ref.~\cite{BeltranJimenez:2019tjy}. 
Its alternative view of gravity also created fertile ground for the proposal of
naturally ghost free modified gravity developments such as $f(T)$ \cite{Ferraro:2006jd,Capozziello:2019cav, Cai_2016}, $f(T,B)$ \cite{Bahamonde:2016grb}, $f(R, T)$ \cite{Bahamonde:2015zma}, or other generalisations such as Lovelock Teleparallel Equivalent  Gravity \cite{Gonzalez:2019tky}, or Conformal TEGR \cite{Maluf:2011kf, Formiga:2019frd,Bamba:2013jqa}.

One of the advantages presented 
by
TEGR over GR is the claim that it can be formulated as a gauge theory for the translation group \cite[see][and references herein]{Aldrovandi:2013wha,Krssak:2018ywd}
; in such formulation
the principal bundle framework of gauge theories is paralleled to extract the TEGR torsion from 
the curvature of a connection defined 
on the principal bundle of translations
. In this approach, the 
so-called canonical one-form, required to define the torsion, is indeed implicitly identified with the connection one-form. Since 
the definition of the canonical one form, as a
structure element
of the frame bundle, does not fit with that of a connection,  this
identification 
becomes problematic
\cite{Fontanini:2018krt,Pereira:2019woq,LeDelliou:2019esi}.
The Cartan connection 
has
therefore been proposed 
as a solution to this difficulty, 
it has in fact the right geometrical properties while being able to offer a bundle description and thus possibly support a gauge theory interpretation.
This paper reviews such proposal, its recent developments, as well as 
additional details on related topics.

Differential geometry definitions and concepts can be referred to in
 \cite{Fecko:2006, Isham:1999qu, Nakahara:2003, KobayashiNomizu:1963}. In Sec.~\ref{SEC-WhichConnec}, we will overview the requirements emerging from TEGR that would be necessary to express it in gauge form. A review of the coherent framework using a Cartan connection for TEGR, with its possible opening to the gauge point of view, is spelled out in Sec.~\ref{SEC-CartanTEGR}, followed, in Sec.~\ref{sec:CartanGR}, by a small note on such framework's application to GR, before concluding in Sec.~\ref{SEC-Conclu}.

\section{Why a Cartan connection to describe TEGR?}\label{SEC-WhichConnec}

\subsection{Constitutive elements 
for gauge-TEGR}\label{SUBSEC-IngredientsForTEGR}

In the context of 
TEGR and gauge theory, let us examine how the constitutive elements 
of TEGR
fit within the framework 
of a typical gauge theory, as found in 
particle physics. 

On physical side, the main constitutive elements 
of such theories are gauge fields and their associated field strength
s
. 
They are dynamical field
s
whose free field equation (uncoupled to matter) exhibits gauge invariance.
They mediate an interaction between matter fields and ensure that the matter field equations are locally 
invariant under some symmetry. On the mathematical side, the gauge fields are sections of Ehresmann connections
defined on a principal $G$-bundle, $G$ being a the global symmetry group of the free matter-field equations. 
The field strengths are (sections of) the curvature of these connection
one-form
s
.

In TEGR, spacetime is a metric manifold $(M, g)$, gravity is carried by torsion, curvature is null, as well as 
non-metricity. 
Taking the Cartan view, the metric is induced by orthonormal (co-)frames (tetrads) through ${\displaystyle \eta(e,e) = g}$, $\eta$ being the metric of Minkowski tangent space. Here we recall 
some central structures related to 
these notions:  
\begin{enumerate}
    \item the  
orthonormal frame bundle $OM$, a principal SO$(1,3)$-bundle, sub-bundle of the frame bundle $FM$: 
its base manifold is the spacetime and its fibers contains all the
orthonormal frames $e$ at a base manifold 
point, each frame being in one-to-one correspondence with an element of SO$(1,3)$ a (local) Lorentz transformation,
    \item the canonical one-form $\theta$, an $\setR^4$-valued one-form, defined through:
\begin{equation}\label{EQ-DefCanno}
 (\theta(e), V) = (e^{-1}, \pi_* V),
\end{equation}
or in coordinates,
\begin{equation*}
 \theta^a(e)[V] = e^a[\pi_*V] = V^a
\end{equation*}
where $e$ is a frame at 
a point $x$ of the base manifold $M$, $e^{-1}$ its co-frame, 
$V$ a vector field of the tangent bundle 
$TOM$, and $\pi$ the projection on the base. The canonical one-form and the co-frames are related through the
important equality
\begin{equation}\label{EQ-link-theta-e}
\sigma^*\theta = e,
\end{equation}
where $\sigma$ is some section of the frame bundle\footnote{Note that, since $\theta$ is an element of structure this relation 
essentially states an equivalence between local sections and local frames. A change of gauge, which in the bundle formalism
of gauge theories is a change of section, here precisely corresponds to a change of frame.}.
\item an Ehresmann connection $\omega$ (a Lorentz or spin connection in this context) defined 
on $OM$ 
allowing us to define the torsion through the usual expression on the frame bundle: 
\begin{equation}\label{EQ-def-torsion}
    \Theta(\omega) = d\theta + \omega\wedge \theta.
\end{equation}

\end{enumerate}

The two objects $OM$ and $\theta$ are defined as soon as a spacetime is present. If in addition 
an Ehresmann connection is defined 
on the frame bundle, these structures allow us to define the torsion through 
Eq.~(\ref{EQ-def-torsion}).
Along some section $\sigma$, this torsion leads to the torsion on the spacetime (base manifold):  
$T = \sigma^*\Theta = de + \widetilde\omega\wedge e$, where 
$\widetilde\omega = \sigma^*\omega$. 

A first observation is that the canonical one-form, the frame bundle (on which it is defined) and a (Lorentz) connection 
are needed to obtain the torsion. By contrast, the curvature, which is defined through the expression: $\Omega(\omega) = d\omega + \omega\wedge\omega$,  can be defined as soon as an Ehresmann connection $\omega$ is defined 
on a principal $G$-bundle
, as
is 
the case for particle physics gauge theor
ies
. A second observation is that in the TEGR framework the curvature of the connection is 
by definition zero, which singles out the Weitzenb\"ock connection: $\omega_{\sss W}$. 

From these considerations, we notice 
that the identification
of the basic objects of TEGR with those of a gauge theory
are far from obvious: although a principal bundle (the frame bundle) is present, the connection naturally associated with 
TEGR 
has a null curvature, and the canonical one-form (entering in the definition of torsion) possesses 
no equivalent in the usual gauge
theoretic framework.

A way out this issue 
is to consider
on physical grounds
that, since in TEGR
gravity manifest
s
itself through torsion, this 
torsion should be considered as the field strength of a yet undetermined gauge theory. A first step in finding a gauge formulation of TEGR can then 
be made by setting 
a connection, defined on the orthonormal frame bundle (or a principal bundle
containing it)\Miv{,} whose torsion is the curvature. 

\subsection{Specific features of gauge TEGR connection}\label{SUBSEC-PossibleConnecTEGR}

If torsion, as defined in Eq.~(\ref{EQ-def-torsion}), needs 
to be the curvature of some connection $\omega_{\sss C}$, 
the sought for
connection  
requires 
to be built out of both the Ehresmann connection $\omega$ appearing in the torsion and the canonical one-form $\theta$.  These two
objects $\omega$ and $\theta$ are different in nature. Let
us briefly recall some of their properties, to finally motivate the 
appearance of a Cartan connection.

The one-form $\omega$ defines an Ehresmann connection in the principal SO$(1,3)$-bundle $OM$. In a general sense,
an Ehresmann connection is a mean to split in a unique way each tangent space
s
of a fiber bundle into vertical 
and horizontal subspaces. As vertical subspaces are always 
defined as tangent spaces to the fibers,\footnote{Note that since the fiber is isomorphic to the group in a Principal G-bundle, the tangent spaces to fibers are all isomorphic to the Lie algebra $\mathfrak{g}$.} the Ehresmann connection uniquely specifies horizontal 
subspaces
and, in this sense, defines uniquely the notion of horizontality in the fiber bundle. 
To be meaningful in a principal $G$-bundle, the splitting between vertical and horizontal subspaces must 
be invariant 
under the action of the symmetry group $G$. This invariant splitting is realized in practice through 
a  
connection one-form $\omega_{\sss E}$ such that: 
\begin{enumerate}
\item it takes its values in the Lie algebra $\mathfrak{g}$ of the Lie group $G$,
\item it satisf
ies
: $R_g^* \omega_{\sss E} = Ad_{g^{-1}}\omega_{\sss E}$, $R_g$ being the right action of $G$ on the bundle,\label{enu:Rg=AdEhresmann}
\item it reduces to the Maurer-Cartan form $\omega_{\sss G}$ of the group $G$ along the fibers: \label{enu:MaurerVertEhresmann}
$\omega_{\sss E}(V) = \omega_{\sss G} (V)$, for any vertical vector $V$. 
\end{enumerate}
Following these properties, at each point of the $G$-bundle, vertical vectors are mapped to the Lie algebra $\mathfrak{g}$ and 
horizontal vectors belong
to the kernel of $\omega_{\sss E}$. 
In addition, 
an Ehresmann 
connection allows one to construct 
parallel transport and its related covariant derivative.\footnote{The relation between these notions
appears more evidently 
in the frame bundle, where 
a parallel-transported frame (so-called autoparallel frame field)  
along a base manifold curve $\tilde \gamma$ 
induces a horizontal curve $\gamma$, i.e. 
whose tangent vector is always horizontal, in the full bundle 
\cite[see for instance][p. 544 19.5.1]{Fecko:2006}} 

The one-form $\theta$ 
Eq.~(\ref{EQ-DefCanno}),  contrary to 
a connection one-form, is canonically and only
defined in the frame bundle (hence on its restriction $OM$). 
The
content 
of Eq.~(\ref{EQ-DefCanno})
can be described as follows: for each frame bundle point, mapped to 
coordinates $(e, x)$
, the action 
of $\theta$, through each of its component
s
$\theta^a$, consists in mapping a bundle's tangent space 
vector $V$ 
to its corresponding component $V^a$ in 
the frame $e$
at the base manifold tangent space point $x$. In comparison with the Ehresmann connection, $\theta$:
\begin{enumerate}
\item 
takes its values in $\setR^4$,
\item 
satisfies: $R_g^* \theta = g^{-1}\theta$, $R_g$ being the right action of the matrix group GL$(4, \setR)$ on $FM$ 
(or SO$(1,3)$  on $OM$),
\item 
is horizontal\footnote{We emphasize that $\theta$ is defined independently from the presence of a connection, which is not responsible here for the horizontality.
} in the sense that: $\theta(V) = 0$, for any vertical vector $V$. 
\end{enumerate}

As is physically 
motivated, 
a connection promotes a global symmetry 
to a local symmetry. At the mathematical level, this local symmetry is encoded in the connection's values 
taken in Lie algebra of the symmetry group
. 
Thus, the orthonormal frame bundle $OM$'s Ehresmann connection $\omega$ takes values in the Lie algebra 
of the Lorentz group. Similarly, the 
TEGR connection $\omega_{\sss C}$ we are seeking, with torsion as field strength, should implement 
local symmetry as well as include 
the additional $\theta$ one-form. Since 
$\theta$ 
takes its values in $\setR^4$, which can be considered 
as the algebra for the translation group, this 
suggests 
 to build an 
$\omega_{\sss C}$ with 
values in the
algebra of the Poincar\'e group  SO$(1,3)\rtimes  \setR^4$. 

The considerations above
would appear 
to point towards a gauge theory with an affine connection defined on the principal Poincar\'e bundle, an affine bundle denoted hereafter $PM$. 
However, 
starting
from the orthonormal frame bundle, it suffices to enlarge its Lorentz connection's algebra, so 
we will consider a more minimalistic generalization: 
a Cartan connection, 
valued in the Poincar\'e algebra
while being defined on the original orthonormal frame bundle $OM$.

\section{Cartan TEGR}\label{SEC-CartanTEGR}

\subsection{The Cartan connection}\label{sec:CartanConnection}
A Cartan connection enters 
in the definition of a Cartan geometry
which
can either be considered to generalise 
Riemann geometry or 
Klein geometry, each themselves offering 
a generalization of 
Euclid 
geometry.\footnote{These generalizations applie\Old{s}, when a metric is present, to positive definite metric (e.g. Riemannian geometry) or to non-positive definite metric (pseudo-Riemannian geometry)
.}
An introduction to Cartan geometries in the context of gravity may be found in \cite{Wise_2010,Catren:2014vza}, while a comprehensive mathematical account is 
given in
\cite{Sharpe:1997}. 
Since 
Riemann geometry is well known in GR, let us briefly discuss the notion of Klein geometry, which is as important as Riemann's,  
for the Cartan generalization. 

\subsubsection{
Klein geometry}\label{sec:KleinGeom}
Instead of introducing space curvature to get away 
from Euclid geometry
, 
as in Riemann geometry
, Klein geometry generalizes 
Euclid spaces to 
homogeneous spaces (also termed maximally symmetric, 
see App. \ref{App-G-space} for a reminder). The latter 
are always 
isomorphic (and consequently identified) to some coset 
space $G/H$ (also termed quotient space), where $G$ is a Lie group and $H$ one of its (closed)
subgroup. This basically defines a Klein geometry $(G, H)$, which is the 
homogeneous space $G/H$. When  $G/H$ is connected, 
a connected Klein geometry 
is obtained
-- 
disconnected homogeneous spaces are irrelevant for Cartan geometry, so we'll focus on connected ones. In this context the familiar three dimensional Euclidean space appears as a particular case of Klein geometry 
where $G =$ SO$(3)\rtimes  \setR^3$ and $H =$ SO$(3)$ the rotation group. 

As they derive from the action of 
Lie groups 
on (differentiable) manifolds (see App.~\ref{App-G-space}), homogeneous spaces are (differentiable) 
manifolds\footnote{Here, the hypothesis that $H$ is a (topologically) closed subgroup of the Lie group $G$ is 
central for the space $G/H$ to be a manifold \cite[see for instance][p. 146]{Sharpe:1997}.}. In four dimensions, Lorentzian homogeneous spaces
are the familiar spaces of:
\begin{itemize}
    \item Minkowski:  $G =$ SO$(1,3)\rtimes  \setR^4$, $H= $ SO$(1,3)$.
    \item de Sitter: $G =$ SO$(1,4)$ $H =$ SO$(1,3)$.
    \item Anti-de Sitter: $G =$ SO$(2,3)$, $H= $ SO$(1,3)$.
\end{itemize}
Many properties of homogeneous spaces 
are inherited from the Lie groups underlying their structures. An important case is the relation 
between their tangent spaces and the Lie algebras of the group G and of its subgoup H: 
the Lie group $G$ can itself be seen as principal fiber bundle of fiber $H$, with 
the homogeneous space $G/H$ 
as base manifold, i.e. 
\sbundle{G}{G/H}{H} \cite[see for instance][p. 55]{KobayashiNomizu:1963}. The 
adjoint representation of 
$H$ (the action of $H$ on its Lie algebra) allows one to associate a vector bundle, with fiber $\mathfrak{g}/\mathfrak{h}$, 
to \sbundle{G}{G/H}{H}. 
This associated bundle can be shown to be isomorphic to 
the tangent bundle of the homogeneous space $G/H$, $T(G/H)$
\cite[see][prop. 5.1 p. 163, for details]{Sharpe:1997}. This important property means, in particular, that each tangent
space to 
a homogeneous space is isomorphic to the quotient $\mathfrak{g}/\mathfrak{h}$, namely that at each point $x$ of $G/H$,
\begin{equation}\label{EQ-TangentIsom-g/h}
   T_x (G/H) \simeq \mathfrak{g}/\mathfrak{h}. 
\end{equation}

In addition to the above relation, 
the tangent spaces inherit, from the aforementioned bundle isomorphism, all the structures defined on
$\mathfrak{g}/\mathfrak{h}$
. In particular a non-degenerate $H$-invariant metric defined on 
$\mathfrak{g}/\mathfrak{h}$ pulls-back to an $H$-invariant metric on the corresponding homogeneous space. A paradigmatic example in GR
is given by the metric of the four dimensional homogeneous (maximally symmetric)  spacetimes of Minkowski, de Sitter and Anti-de Sitter \cite{Wise_2010}, inherited from the symmetries of the corresponding coset spaces.

\subsubsection{Cartan geometry}\label{SUBSUBSEC-CartanGeom}
We introduced Sec.~\ref{sec:CartanConnection} recalling that Cartan geometry represents 
at the same time a generalization of Klein and 
Riemann geometries. Here we will give more 
details
on how it actually does merge 
both. In Sec.~\ref{sec:KleinGeom}, we discussed how 
Klein geometry generalizes Euclid 
by considering the Euclidean space
as a particular case of homogeneous spaces, and how these spaces are defined by the quotient of a Lie group $G$ by one of its (closed) subgroups $H$, 
sharing 
the symmetries of the group $G$. Riemann geometry is historically an $N$-dimensional generalisation of the two-dimensional Gaussian differential 
geometry of surfaces departing from the Euclid (plane) geometry by introducing curvature. In contemporary terminology Riemann geometries are 
metric orientable differentiable manifolds, equipped with the Levi-Civita connection
, their
curvature measures how they locally depart from
the flatness of their Euclidean tangent spaces, while the connection relates tangent spaces at different points.

In Cartan geometry 
the
Euclidean tangent space is generalized to a tangent homogeneous space. To simplify visualisation of such tangent spaces, the two-dimensional case can be examined: 
the Riemann view 
considers a tangent plane at each point of the surface, its contact point with the surface constituting its origin, when identified with
the (invariant) vector space $\setR^2$ (under the rotation group SO$(2)$).  By contrast, the  
Cartan view rolls, without slipping, a unique space on the surface
, a
space 
which
can either be a plane, sphere or hyperbolo\"id, one of the three homogeneous space
s
in two dimensions. The contact point thus moves in the homogeneous space, putting in evidence its affine nature. The case of a plane therefore identifies with the affine plane $\setR^2$, invariant under the affine group SO$(2)\rtimes  \setR^2$.

These two different descriptions of tangent spaces are reflected respectively in
the Cartan and Levi-Civita (generalised to affine Ehresmann) connections.
They both provide methods of transfering objects between points joined by some
(continuous) path $\gamma$. The Cartan connection effects the transfer of contact point, rolling without
slipping, in the homogeneous space, while the Levi-Civita connection provides a rule to transfer objects along the path $\gamma$ from one plane to another.

The defining properties of a Cartan geometry are gathered in the following definitions:
\begin{itemize}
    \item \emph{A Cartan geometry}, modeled over a Klein geometry (G, H), is a principal $H$-bundle $\sbundle{P}{M}{H}$ equipped with a Cartan
    connection.
    \item \emph{A Cartan connection} is defined through a one-form $\omega_{\sss C}$ such that:
\begin{enumerate}
\item it takes values in the algebra $\mathfrak{g}\supset\mathfrak{h}$ of $G \supset H$.\label{enu:Gval}
\item it is, at each point $p$ of the $H$-bundle, a linear isomorphism between the tangent space $T_pP$ at $p$ and the 
Lie algebra $\mathfrak{g}$. 
This property requires that $G$ and the tangent space $T_p P$ share 
the same dimension
.\label{enu:gConn}
\item it satisfies: $R_h^* \omega_{\sss C} = Ad_{h^{-1}}\omega_{\sss C}$, $R_h$ being the right action of $H$ on the bundle,\label{enu:Rg=Ad}
\item it reduces to the Maurer-Cartan form $\omega_{\sss H}$ of the group $H$ along the fibers.\label{enu:MaurerVert}
\end{enumerate}
\end{itemize}

The Cartan geometry definition 
sets its 
own
mathematical framework
. In the TEGR situation we are interested in, 
the $H$-bundle will be the orthonormal frame bundle $OM$ with $H=$ SO$(1,3)$ and $M$, the spacetime. In accordance with
the conclusion of Sec.~\ref{SUBSEC-PossibleConnecTEGR}, we will take, for the group $G$, 
the Poincar\'e group and, consequently, for 
the homogeneous space (the Klein geometry), we will use 
the Minkowski space.

The case we are interested in then
presents 
an additional property
: 
its
Cartan geometry is reductive. 
Formally \cite[see][p. 197, for details]{Sharpe:1997}, 
this
means that the Lie algebra decomposition as vector space $\mathfrak{g} = \mathfrak{h} \oplus \mathfrak{g/h}$ is  Ad($H)$-invariant. In practice
, among other results, it implies that 
any $\mathfrak{g}$-valued form defined on $P$ splits along this decomposition 
and, in
particular, the connection 
one-form $\omega_{\sss C}$ splits into 
$\mathfrak{h}$  and 
$\mathfrak{g/h}$ parts.
For the Poincar\'e algebra, 
these correspond respectively to the 
Lorentz 
and 
translation parts
, $\mathfrak{h} = \mathfrak{so}(1,3)$ and 
$\mathfrak{g/h} = \setR^4$. 
Consequently, the 
reductive Cartan connection one-form 
splits
into
\begin{equation}\label{EQ-CartanOneFormReductive}
 \omega_{\sss C} = \omega + \theta,
\end{equation}
where $\omega$ is an 
$\mathfrak{h}$-valued Ehresmann type connection one-form
,  
and $\theta$ a 
$\mathfrak{g}/\mathfrak{h}$-valued one-form, both defined on the principal $H$-bundle.
This decomposition also follows for 
the curvature of $\omega_{\sss C}$,
\begin{equation}\label{EQ-CartanCurvatureReductive}
\Omega(\omega_{\sss C}) = \Omega(\omega) + \Theta(\omega),
\end{equation}
where the curvature of the $\omega$ part reads $\Omega(\omega) = d\omega + \omega \wedge \omega$ 
and both parts combine in the torsion $\Theta(\omega) = d\theta + \omega\wedge \theta$
.

The 
definition of the Cartan connection, 
through 
$\omega_{\sss C}$, differs from that of the 
Ehresmann connection, in Sec.~\ref{SUBSEC-PossibleConnecTEGR}, in the first two 
properties 
which we will now discuss
. 

From 
property \ref{enu:Gval}, instead of being valued in the Lie algebra of the fiber $\mathfrak{h}$, 
$\omega_{\sss C}$ is valued in the whole $\mathfrak{g}$ algebra
. 
This agrees 
with the ``minimal extension" mentioned 
at the end of Sec.~\ref{SUBSEC-PossibleConnecTEGR}
since, of the full algebra $\mathfrak{g}$, only appears the translations part $\mathfrak{g/h}$.

Property \ref{enu:gConn} relies both on the Klein geometry and the soldering property. It imposes the tangent space, $T_pP$ of $\pbundle{P}{M}{H}{\pi}$ at $p$, and $\mathfrak{g}$ to be isomorphic.
In addition, its vertical part is always defined as the tangent space to the fiber and is also isomorphic to $\mathfrak{h}$; Eq.~(\ref{EQ-TangentIsom-g/h}) implies that the orthogonal 
complement $\mathfrak{g/h}$ of $\mathfrak{h}$ in the decomposition $\mathfrak{g} = \mathfrak{h} \oplus \mathfrak{g/h}$ identifies with the tangent space $T_{\pi(p)}(G/H)$ to the homogeneous space at $\pi(p)$; therefore, it can be identified with the tangent space $T_{\pi(p)}(M)$ to the base manifold at $\pi(p)$. In the same way as $T_{\pi(p)}(M)$ is a fiber of the tangent bundle $TM$, $T_{\pi(p)}(G/H)$ 
is the fiber of a ``tangent homogeneous-space" bundle of $M$ associated to $P$: $P\times_{\sss H} G/H$. Since the connection induces $\mathfrak{g/h}\simeq T_{\pi(p)}(G/H)\simeq T_{\pi(p)}(M)$, that identifies a tangent space of $M$ to a tangent
space of a fiber of  $P\times_{\sss H} G/H$, it thus effects a soldering \cite[see for instance][App.~3]{Fontanini:2018krt}
of the $H$-principal bundle to its base $M$, illustrating this ``built-in property'' of the Cartan connection.

Finally, properties \ref{enu:Rg=Ad} and \ref{enu:MaurerVert} correspond identically to properties \ref{enu:Rg=AdEhresmann} and \ref{enu:MaurerVertEhresmann} for Ehresmann connection one-forms.

Returning to the frame bundle we consider for TEGR (with its Cartan connection), and as anticipated in Eq.~(\ref{EQ-CartanOneFormReductive}) notation, its canonical one-form $\theta$ does indeed correspond to the 
$\mathfrak{g/h}$ term in the Cartan connection.
The interpretation of the Cartan geometry as a movement of the tangent space  
rolling without slipping on the base, added with Eq.~(\ref{EQ-link-theta-e}), allows us to interpret that term of
the Cartan connection in link with the so-called moving (co)frame.

\subsubsection{The TEGR connection}\label{SUBSUBSEC-TEGRConnection}
From the discussion above and the requirement that for TEGR
the curvature of the connection yields 
the torsion,  
Eq.~(\ref{EQ-CartanOneFormReductive}) specializes to
\begin{equation}\label{EQ-CartanReducConnec-W+thet}
 \omega_{\sss C} = \omega_{\sss W} + \theta,   
\end{equation} 
where the Ehressman term, 
$\omega_{\sss W}$,  stands for 
the curvature-less Weitzenb\"ock  
connection  
and where 
$\theta$ coincides with the canonical one-form on $OM$. Then, through 
Eq.~(\ref{EQ-CartanCurvatureReductive}) or by 
direct calculation \cite[see][]{LeDelliou:2019esi}, the required relation
$\Omega(\omega_{\sss C}) = \Theta(\omega_{\sss W})$ is obtained.

As we previously pointed out 
while analyzing
their differences, 
the Cartan connection
, that we suggest describes TEGR, contrasts with 
the Ehresmann  connection
, that represents 
particle physics gauge theories. At this stage
it is premature to claim whether this framework allows TEGR to 
be seen as a gauge theory
, however
, in the Cartan geometry, 
$\omega_{\sss C}$ does give 
the expected field strength, in addition to relating gravitation to 
the translation symmetry
, in accordance with Noether's theorem: while 
each term of $\omega_{\sss C}$ in Eq.~(\ref{EQ-CartanReducConnec-W+thet}) corresponds to Lorentz and translation symmetries
respectively 
-- through values in the 
Poincar\'e Lie algebra  --  and as the Weitzenb\"ock curvature vanishes, the only contributing 
curvature (field strength){ 
term} comes from the translation 
($\mathfrak{g/h} = \setR^4$) valued $\theta$ 
term. 
In 
contrast with the difficulties to extract a translation term from 
the usual formulation of GR (see App.~\ref{APP-TranslationAndGauge}), the natural appearance of translation with $\theta$ in Cartan 
TEGR 
counts as its achievement. Furthermore, to preserve the physical 
relation between field strength and 
gauge field 
in the 
Cartan TEGR framework, Eq.~(\ref{EQ-link-theta-e}) points 
toward identifying 
the co-frame as the 
gauge field of the theory. To settle the identification 
of a gauge field in the theory requires 
to examine the coupling to matter, 
since 
particle physics coupling 
is mediated by the gauge field.

\subsection{Matter coupling with Cartan connection}\label{SEC-CoupMatDetails}

In particle physics gauge theories, charged matter fields interact through the exchange of gauge bosons, mediating the interaction.  At the 
classical 
level, this interaction is described thanks to a covariant derivative ensuring the minimal coupling. That covariant derivative is directly
related to an Ehresmann connection 
whose
pullback on the base manifold (the Minkowski space) is the gauge field. In fact, covariant 
derivative and
parallel transport are directly 
built from an  
Ehresmann connection in a principal bundle. 
In classical GR the coupling to classical matter field is also realized through a covariant derivative: 
in Cartan (tetrads) 
formalism, 
the covariant
derivative is implemented from 
the Levi-Civita connection one-form $\omega_{\sss LC}$
in the frame bundle. 
So far
, GR -- and thus the Levi-Civita connection -- 
reproduces dramatically well 
observational 
data\footnote{Unfortunately, the interpretation of the Levi-Civita connection as a  gauge field mediating gravitation 
is not satisfying (see App.~\ref{APP-TranslationAndGauge}).} and hence compels any 
formulation of gravity 
to reproduce the Levi-Civita coupling.

\subsubsection{From Cartan connection to Levi-Civita coupling}\label{SUBSUBSEC-CartanToLeviCivita}

In this section we will recall from \cite{huguet2020teleparallel} how to obtain the Levi-Civita part of coupling to matter from the Cartan connection one-form $\omega_{\sss C}$, proposed to describe TEGR. 

Starting from the Cartan geometry, the 
Levi-Civita coupling is obtained in three main steps
, commented hereafter
\begin{enumerate}
\item The Cartan connection one-form $\omega_{\sss C}$ on $OM$ is mapped 
to an Ehresmann connection one-form $\omega_{\sss E}$ thanks 
to the Sharpe theorem \cite[Prop. 3.1 p. 365 of][]{Sharpe:1997} (also reproduced in \cite{huguet2020teleparallel}).
\item The  Weitznb\"ock term, present in $\omega_{\sss E}$: the first term of the r.h.s. of Eq. (\ref{EQ-CartanReducConnec-W+thet}), 
is mapped to 
the Levi-Civita one-form thanks to 
the contorsion one-form  defined on $OM$ \cite[theorem 6.2.5 p. 79]{Bleecker:1981}.
\item The resulting one-form, an affine connection on the Poincar\'e principal bundle  $PM$ , is then pulled back on $OM$  with 
a 
map \cite[proposition 3.1 p. 127]{KobayashiNomizu:1963} that splits it 
into 
the canonical and the Levi-Civita one-forms, 
parallel transport 
and covariant derivative proceeding then from the Levi-Civita Ehresmann connection, now in a Cartan (tetrads) formalism setting.
\end{enumerate}

More explicitly: as the Cartan connection does not define canonically parallel transport
, by contrast with Ehresmann connection, the first step relates, in our peculiar case, the set of 
$\omega_{\sss C}$  
to the set of affine Ehresmann connections 
on the 
affine (Poincar\'e) bundle,  as obtained
in \cite{huguet2020teleparallel} with the help of Sharpe theorem\footnote{A 
technical condition restricts 
the set of possible Ehresmann connections. However, 
since it 
bears
no influence on 
our results, we just point this out to the interested reader.}. 
As the theorem, in our case, results in the identity of the 
connection one-form 
on the base manifold (the spacetime), the 
pullback of the 
Weitzenb\"ock term $\omega_{\sss W}$ in 
Eq.~(\ref{EQ-CartanReducConnec-W+thet})
is still present. 
To obtain the correct matter coupling requires to map this 
term 
to the Levi-Civita one-form $\omega_{\sss LC}$.
 
This is done, in the second step,
applying the theorem\footnote{The proof in the bundle formalism is given in  \cite[theorem 6.2.5 p. 79]{Bleecker:1981}.} that relates 
any Ehresmann one-form from 
the orthonormal frame bundle $OM$ to the 
Levi-Civita one-form via its contorsion one-form
. 

Two remarks are in order here. First, although 
any 
two Ehresmann connections are always related by some specific 
one-form
, the existence of 
the contorsion, and hence getting the Levi-Civita one-form, relies on the existence of 
a metric
. That metric is present in $OM$, where the 
possibility to map Weitzenb\"ock to Levi-Civita one-forms can 
be traced back to the choice of the 
Minkowski metric for tangent spaces to the base (the spacetime), and in this sense related to the Equivalence Principle. 
The second remark concerns the building of the Levi-Civita one-form in this context: as 
contorsion is composed from 
torsion, itself built out 
of the Weitzenb\"ock 
connection, and of the canonical one-form, 
the resulting 
Levi-Civita one-form 
is 
a function of the Weitzenb\"ock one-form%
, the canonical one-form, and its
derivatives.

This Weitzenb\"ock-to-Levi-Civita 
mapping 
allows us to redefine the Ehresmann one-form $\omega_{\sss E}$, into 
$\widetilde\omega_{\sss E}$, 
now the sum of a linear, 
Levi-Civita one-form, $\omega_{\sss LC}$, and a translation term, 
the canonical one form 
$\theta$, also 
defined on the affine bundle $PM$. The third step maps back 
$\widetilde\omega_{\sss E}$ into $OM$ 
\cite[proposition 3.1 p. 127]{KobayashiNomizu:1963}, where it splits into $\omega_{\sss LC}$, and $\theta$
.

Compositing these steps, the final result yields a 
map on 
$OM$ 
from 
Cartan to 
Levi-Civita connection
:
\begin{equation}\label{EQ-mapCartanLeviCivita}
    \omega_{\sss C} \longmapsto \omega_{\sss E} = \omega_{\sss W} + \theta 
    \longmapsto  \widetilde\omega_{\sss E} = \omega_{\sss LC} + \theta \longmapsto (\omega_{\sss LC},  \theta).
\end{equation}
This 
map 
provides 
matter coupling 
to TEGR-gravity in agreement with 
the familiar coupling of GR, as well as 
the Fock-Ivanenko derivative in Cartan (tetrads) formalism. 

\subsubsection{A remark about the gravitational field in Cartan TEGR}\label{SEUBSUBSEC-RemarkGravField}


In classical GR, the gravitational interaction is carried by the metric $g$ or, in Cartan (tetrads) formalism, by the frame field $e$. 
The metric, or the frame field, also entirely determines the Levi-Civita connection.
Although the Cartan TEGR framework reaches these results from an entirely different perspective, $e$ remains the dynamical field and coupling remains 
mediated through the Levi-Civita connection. The gravitational field thus remains carried by the frame field in Cartan TEGR.

\subsection{Is Cartan TEGR a gauge theory?}
As discussed Sec.~\ref{SUBSEC-IngredientsForTEGR}, TEGR requires some crucial structural elements that 
are absent from 
particle physics gauge theories. Therefore, 
adopting the usual particle physics structure as gauge theory 
definition 
would 
exclude TEGR
. In the wider context of gauge-gravity theories \cite[see][for a comprehensive, 
historical account]{Blagojevic:2013xpa} some rather deep change of paradigm were considered 
\cite{hehl2020conservation}. Here, following \cite{huguet2020teleparallel} on physical grounds,  we only minimally propose to extend 
the usual gauge theory framework 
to include the required extra structure for 
TEGR
. 
Although this extension hints at a possible interpretation of Cartan TEGR as a gauge theory, which we discuss below, we do not claim adhesion to it, leaving interpretation open.

\subsubsection*{Frames as gauge fields}

The 
Cartan connection $\omega_{\sss C}$, as decomposed in 
Eq.~(\ref{EQ-CartanReducConnec-W+thet}), yields a curvature 
equal to the torsion. Moreover, since  
the curvature of the Weitzenb\"ock connection vanishes, the only contributing dynamical field to the field strength remains the field of 
frames. This, as pointed out in Sec.~\ref{SUBSUBSEC-TEGRConnection}, hints at the field of frames behaving as a gauge field. 
We also pointed out 
Sec.~\ref{SEUBSUBSEC-RemarkGravField} 
that the frame can be considered as mediating the gravitational field. This is already the case 
in GR  
since 
the Levi-Civita connection, determined from the metric, thus the frames, in a 
Cartan (tetrads) formulation, mediates coupling to tensorial and spinorial fields. Additionally, since 
classical scalar fields also couple to 
gravity, while their covariant derivative reduces 
to a partial derivative, 
coupling should be present in the partial derivative. As 
the tangent vector $\partial_\mu$ decomposes in the frame basis $e_a$ following 
$\partial_\mu = (\partial_\mu)^a e_a$, such coupling reinforces the interpretation of the frame field as mediating 
gravitational interaction. 

All the 
above arguments advocate for 
the interpretation of
$e$ as the gauge field 
on the basis that it mediates 
interaction.
At the same time, interpreting  
$e$ as a gauge field 
departs from 
the paradigmatic framework of gauge theories, that would relate $e$ to an Ehresmann 
connection by a simple pullback on the base manifold: 
it relates instead to the 
Cartan connection 
$\omega_{\sss C}$ through the composite map combining Eqs. (\ref{EQ-mapCartanLeviCivita}) and 
(\ref{EQ-link-theta-e}). To interpret TEGR as a gauge theory thus requires to abandon the strict 
correspondence between
gauge field and 
Ehresmann connection
. 
 Note that the usual particle physics gauge theories remain unaffected by this change
 , 
 since it 
 only involves external ingredients.


In addition to 
mediating interaction, particle physics gauge fields also implement some local (i.e. 
spacetime point dependent) 
symmetry invariance in matter field equations. As 
a frame $e$, from Eq.~(\ref{EQ-link-theta-e}), is clearly related to 
the canonical one-form $\theta$, 
valued in the translation part
of the Poincar\'e algebra, it thus relates to local and infinitesimal translations
. The discussion above reveals that matter coupling involves two occurrences of the frame $e$: in the expansion of 
the partial derivative 
and in the 
Levi-Civita connection, seen  as 
the difference between 
the non-dynamical Weitzenb\"ock connection and the frame "induced" contorsion
. 
However, the field $e$ does not emerge from localisation of a global translation symmetry. Instead, the symmetry is already local and infinitesimal since $e$ only relates to the translation algebra.
The 
usual process involved in 
"gauging a theory" by 
which a 
globally invariant matter field 
Lagrangian becomes 
locally invariant with the introduction of the gauge field in the derivatives is not present in this context. 
Moreover, the basic symmetry of GR, i.e. diffeomorphism invariance (the invariance under spacetime 
coordinate 
changes), does not proceed from translation symmetry but is already 
a built-in property of 
differential geometry.

From 
the above discussion, 
the frame $e$, 
on the basis that it mediates the interaction with matter fields, can be perceived to share physical properties of 
a gauge field. Nevertheless, such an interpretation
requires to enlarge 
the notion of gauge field 
compared with its 
meaning 
in usual particle field theories. 
In particular, the frame field $e$ is 
the pullback of 
the canonical one-form, rather than that of an Ehresmann connection, and is
the translation part of a Cartan connection 
whose
curvature, still considered as the field strength, is the torsion.    

\section{A note on Cartan GR}\label{sec:CartanGR}
As TEGR is equivalent to GR, the Cartan TEGR structure to represent TEGR can be transposed to use a Cartan connection to describe GR. This transposition can be obtained by
\begin{enumerate}
    \item setting the Ehresmann term
    in the Cartan connection one-form (\ref{EQ-CartanOneFormReductive}) to the Levi-Civita connection: $\omega_{\sss C} = \omega_{\sss LC} + \theta$,
    \item  recognising the mapping using contortion reduces to identity, as torsion -- and thus contorsion -- vanishes for the Levi-Civita connection, such that the  map (\ref{EQ-mapCartanLeviCivita}) becomes
    \begin{equation*}
    \omega_{\sss C} \longmapsto  \omega_{\sss E} = \omega_{\sss LC} + \theta \longmapsto (\omega_{\sss LC},  \theta).
\end{equation*}
\end{enumerate}
The Cartan curvature of the corresponding $\omega_{\sss C}$ then reduces to the Riemann curvature term of the Levi-Civita connection, as expected in GR.
Note that the parallel transport for such reductive Cartan connection could be directly seen as induced, in the initial bundle, by the Ehresmann part of its one-form $\omega_{\sss C}$. The only difference of this framework with the usual Cartan (tetrad) formalism of GR lies in the Cartan connection and the possibility it awards GR, in a similar way as for TEGR above, to be conceivably interpreted as a gauge theory. Similar modifications to particle physics type gauge theory are required, in the lines described above.
The possible interpretation of the frame field  as a gauge field for translations could follow similar arguments 
to the ones 
presented
above, leading to a recognition of the resulting Cartan GR as a gauge theory to be similarly left open.

We note that the use of Cartan connection to describe GR has already been studied, along other lines, in \cite{Catren:2014vza, jordan2014gauge}.

\section{Conclusion}\label{SEC-Conclu}

In this paper we have reviewed our proposal  formulating TEGR from a Cartan type connection. The theory we obtain, and refer to as Cartan TEGR, produces 
identical predictions 
to TEGR and thus to GR. It provides a consistent mathematical framework in which the torsion is
obtained as the curvature of a reductive Cartan connection. This connection is the sum of 
the curvatureless Weitzenb\"ock connection and the 
so-called canonical one-form 
whose
pullback along some section on spacetime is a frame. The usual (GR) coupling to matter fields is obtained thanks to Sharpe's theorem relating Cartan and Ehresmann connections, and to
the introduction of the contorsion one-form in relation with the equivalence principle. 

Cartan TEGR is then examined from a gauge theoretic perspective. This is done without claiming it is a gauge theory -- this interpretation is left open -- and we discuss the mandatory extensions and the departures from the particle physics gauge theory paradigm. This perspective allows us to specify 
to which extent the (local) field of frame could be recognized as a translation gauge field 
mediating gravity, and how it could be 
associated to torsion, 
as a field strength. 

%

\section*{Acknowledgements}

The authors wish to thanks D.~Bennequin and T.~Lawrence for helpful discussions, respectively, on geometry and on the status of diffeomorphisms in groups and gravity. The work of M.~Le~D. has been supported by 
Lanzhou University starting fund, and the Fundamental Research Funds for the
Central Universities (Grant No.lzujbky-2019-25).

\medskip
\appendix
\section{Right (left) G-space, homogeneous space and coset space G/H}\label{App-G-space}
We follow Fecko \cite{Fecko:2006} (Sec. 13.1, 13.2 for details). A manifold $M$ on which a right (left) action of a Lie group G is available is a 
right (left) G-space. The orbit of some point $x$ in 
$M$ is the set of points which can be reached from $x$ by the group action. Note 
that orbits are disjoints. In the special case when $M$ only contains 
a single orbit, that is when the action of G is transitive on $M$, 
the G-space is called a homogeneous space.\footnote{G transitive on $M$ means there exists for all pair of point (x, y) in $M$ $g\in$ G such that $R_g x = y$ ($L_g x =y$).}

The left (right) coset space G/H, where H is a topologically closed subgroup of G, is the set of equivalent classes $[g]$ for the 
relation noted\footnote{The map $\sim$ is defined by:  
$g_1\sim g_2$ iff 
there exists an 
$h\in$ H such that $g_1 = g_2 h$ ($g_1 = h g_2$).}
$\sim$.
Since 
the left product in G $gg'$ is transitive, 
the left action in G/H defined through
$L_g [g'] := [gg']$ is then transitive, and consequently G/H is a homogeneous space.

In this context
, the group G does act, through $L_g$, on G/H, and therefore 
this coset space can be considered as 
a left G-space. In fact, one can show that all homogeneous
G-spaces are obtained in this way, which means that any 
homogeneous space manifold $M$ with a group action from 
G 
is isomorphic to a G/H for 
some closed subset H. 

Moreover, one can show that H  can be taken as the stabilizer of some point $x$ of $M$.\footnote{The stabilizer of some point $x$ of $M$ with a group action from G is the subset G$_x$ of G 
(which is indeed a subgroup) whose elements leave $x$ invariant.} Since M is homogeneous, any two points being in the same orbit share isomorphic stabilizers and the choice of the 
point $x$ is un
important
.

Finally, we note that G is a principal H-bundle over the (left) coset space G/H if G is a Lie group and  H, 
a (topologically) closed (not necessarily invariant) subgroup 
 \cite[see for instance][Exemple 5.1 p.66]{KobayashiNomizu:1963}. 
Here the action of H on G is just the right multiplication. The fibers are the left cosets 
of H and since the fiber containing the identity 
is naturally isomorphic to H, all fibers are isomorphic to H.\linebreak

\section{A note on gravity as a gauge theory of translations: a not so obvious relation}\label{APP-TranslationAndGauge}
Since Noether theorem associates translations to the stress-energy tensor, it 
suggests that gravity, as described by the Einstein equations
of General Relativity (GR), is related to translations.  Whereas translation is a global symmetry of physical equations in Minkowski 
space, it can become 
local 
when gravity is present. This, in turn, open 
the possibility that gravity arises from a gauge theory of translations, where, following the usual gauge theory construction
, 
a gauge field is minimally coupled to the
(Minkowskian) matter fields in order to obtain a matter equation locally invariant under translations.\footnote{A 
point of terminology: by translation we mean an element of the translation group $(\setR^4, +)$, often abbreviated in $\setR^4$. The translations, 
even local, should not be confused, in GR, with the ``local displacements" generated by the group of local diffeomorphisms, or more 
precisely, its Lie algebra, sometimes also referred as to local translations.} 

Unfortunately, the situation is more complicated. A first difficulty concerns 
the structure of a gauge theory: contrary to the action of particle physics gauge groups, involving 
matter and 
gauge fields, translations also act on space-time (
often termed  
``external" symmetry). This relates to 
the soldering property \cite[see for instance][App.~3]{Fontanini:2018krt}. 
Since gauge 
theories of particle physics' 
standard model 
are built on internal groups (U$(1)$, SU$(2)$, \emph{etc.}), they are not concerned with 
this property. On the contrary, 
a translation gauge theory should take soldering 
into account
. 
This suggests to enlarge 
the mathematical framework of gauge-theories to include this ``new" ingredient.
A second difficulty stems from 
the symmetries under consideration: 
in gauge field theory the local invariance is implemented thanks to the connection (valued in the Lie algebra of the symmetry group) 
appearing in the covariant derivative. In general relativity, 
in the Cartan (tetrads) formalism, 
matter coupling 
involves the so-called spin or Lorentz connection. As 
it takes its value in the Lorentz algebra $\frak{so}(1,3)$, this connection takes its source in the Lorentz group SO$(1,3)$ and so isn't related to the translation group.

As a result, the encoding of the natural link made by Noether theorem between gravity an translation symmetry into a gauge theory
is not a straightforward issue. 

\bibliography{Cartan-TEGR-Review}
\end{document}